\documentclass{article}

\usepackage{arxiv}

\usepackage[utf8]{inputenc} 
\usepackage[T1]{fontenc}    
\usepackage{hyperref}       
\usepackage{url}            
\usepackage{booktabs}       
\usepackage{amsfonts}       
\usepackage{nicefrac}       
\usepackage{microtype}      
\usepackage{lipsum}		
\usepackage{graphicx}
\usepackage{doi}
\usepackage{multirow}
\usepackage{float}

\title{Ensembling and Test Augmentation for Covid-19 Detection and Covid-19 Domain Adaptation from 3D CT-Scans}

\author{ \href{https://sciprofiles.com/profile/FaresBougourzi}{\includegraphics[scale=0.06]{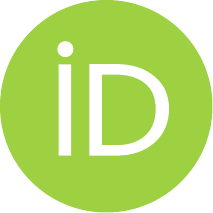}\hspace{1mm}Fares BOUGOURZI}
\thanks{.} \\
Junia, UMR 8520, CNRS, Centrale Lille, Univerity of Polytechnique Hauts-de-France, 59000 Lille, France \\
	\texttt{faresbougourzi@gmail.com; fares.bougourzi@junia.com} \\
	\And
	\href{https://orcid.org/0000-0001-5077-4862}{\includegraphics[scale=0.06]{orcid.pdf}\hspace{1mm}Feryal WINDAL MOULAÏ} \\
Junia, UMR 8520, CNRS, Centrale Lille, Univerity of \\ Polytechnique Hauts-de-France, 59000 Lille, France \\
	\texttt{feryal.windal@junia.com} \\ 
	\And
	\href{https://orcid.org/0000-0000-0000-0000}{\includegraphics[scale=0.06]{orcid.pdf}\hspace{1mm}Halim Benhabiles} \\
	Centre d' Enseignement, de Recherche et \\
   d'Innovation Systèmes Numériques, Lille, France \\
	\texttt{halim.benhabiles@imt-nord-europe.fr} \\ 
	\And
	\href{https://orcid.org/0000-0001-6581-9680}{\includegraphics[scale=0.06]{orcid.pdf}\hspace{1mm}Fadi DORNAIKA} \\
	University of the Basque Country UPV/EHU,\\
	San Sebastian, SPAIN; IKERBASQUE, Basque \\ Foundation for Science, Bilbao, SPAIN \\
	\texttt{fadi.dornaika@ehu.eus} \\ 
	\And
	\href{https://orcid.org/0000-0001-7218-3799}{\includegraphics[scale=0.06]{orcid.pdf}\hspace{1mm}Abdelmalik Taleb-Ahmed} \\
	IEMN UMR CNRS 8520, Université \\Polytechnique Hauts de France, UPHF\\
	\texttt{Abdelmalik.Taleb-Ahmed@uphf.fr} \\
}

\renewcommand{\shorttitle}{\textit{arXiv} Template}

\hypersetup{
pdftitle={A template for the arxiv style},
pdfsubject={q-bio.NC, q-bio.QM},
pdfauthor={David S.~Hippocampus, Elias D.~Striatum},
pdfkeywords={First keyword, Second keyword, More},
}

\begin{document}
\maketitle

\begin{abstract}

Since the emergence of Covid-19 in late 2019, medical image analysis using artificial intelligence (AI) has emerged as a crucial research area, particularly with the utility of CT-scan imaging for disease diagnosis. This paper contributes to the 4th COV19D competition, focusing on Covid-19 Detection and Covid-19 Domain Adaptation Challenges. Our approach centers on lung segmentation and Covid-19 infection segmentation employing the recent CNN-based segmentation architecture PDAtt-Unet, which simultaneously segments lung regions and infections. Departing from traditional methods, we concatenate the input slice (grayscale) with segmented lung and infection, generating three input channels akin to color channels. Additionally, we employ three 3D CNN backbones—Customized Hybrid-DeCoVNet, along with pretrained 3D-Resnet-18 and 3D-Resnet-50 models—to train Covid-19 recognition for both challenges. Furthermore, we explore ensemble approaches and testing augmentation to enhance performance. Comparison with baseline results underscores the substantial efficiency of our approach, with a significant margin in terms of F1-score (14\%). This study advances the field by presenting a comprehensive methodology for accurate Covid-19 detection and adaptation, leveraging cutting-edge AI techniques in medical image analysis.

\end{abstract}

\keywords{Covid-19 \and Deep Leaning \and CNNs \and Recognition}

\section{Introduction}
Since the onset of the Covid-19 pandemic in late 2019, Reverse Transcription-Polymerase Chain Reaction (RT-PCR) has been widely established as the primary method for Covid-19 detection. Nevertheless, this testing modality presents several drawbacks, including limited availability of RT-PCR kits, lengthy procedures, and a notable incidence of false negative results \cite{jin_rapid_2020}. Consequently, medical imaging techniques such as X-rays and CT-scans have gained prominence as complementary tools for Covid-19 detection \cite{bougourzi_recognition_2021,bougourzi_emb-trattunet_2024}. CT-scans not only serve in identifying Covid-19 infections but also play a crucial role in monitoring patients' conditions and predicting disease severity \cite{bougourzi_emb-trattunet_2024,bougourzi_challenge_2021}.

In recent years, Deep Learning methodologies have risen to prominence in computer vision tasks, showcasing remarkable performance gains over conventional techniques \cite{bougourzi_fusing_2020, bougourzi_deep_2022, bougourzi2023cnn}. However, one of the primary challenges associated with Deep Learning, particularly in the realm of Convolutional Neural Networks (CNNs), lies in the necessity for extensive labeled datasets, a resource often scarce in medical domains \cite{bougourzi_emb-trattunet_2024, bougourzi2023pdatt}. Furthermore, the majority of existing CNN architectures are tailored for processing static images, which proves inadequate in capturing the intricacies inherent in medical imaging data, especially for the volumetric scans \cite{bougourzi_challenge_2021}. On the other hand, domain adaptation is one of the most challenging aspects encountered in medical imaging, owing to the high variability of data from one center to another due to the variety of recording settings and scanners. Machine learning techniques used in computer-aided medical image analysis usually suffer from the domain shift problem caused by different distributions between source/reference data and target data \cite{guan2021domain}.

In our paper, we present an  approach to address both Covid-19 detection and domain adaptation challenges on the 4th COV19D competition. Our method revolves around lung segmentation and Covid-19 infection segmentation using the PDAtt-Unet CNN-based segmentation architecture, which concurrently segments lung regions and infections. Departing from traditional methods, we integrate the input slice with segmented lung and infection, creating three input channels akin to color channels. We utilize three 3D CNN backbones— Customized Hybrid-DeCoVNet, pretrained 3D-Resnet-18, and 3D-Resnet-50 models—to train Covid-19 recognition for both challenges. Additionally, we explore ensemble approaches and testing augmentation to enhance performance. 
Our main contributions are:

\begin{itemize} 
\item We adopted a Customized Hybrid-DeCoVNet architecture for both Covid-19 Detection and Covid-19 Domain Adaptation Challenges. This architecture incorporates the concatenation of the original slice, the segmented lung, and the segmented Covid-19 infection as the three input channels.

\item In addition to our proposed Customized Hybrid-DeCoVNet architecture, we leveraged two pretrained 3D-CNNs: 3D-Resnet-18 and 3D-Resnet-50.

\item We explored ensemble approaches and testing augmentation techniques to enhance the robustness and performance of our method.

\item Our approach demonstrated a substantial improvement in efficiency compared to baseline results, with a significant margin in F1-score (14\%).

\item We have made our codes and pretrained models publicly available in \footnote{\url{https://github.com/faresbougourzi/4th-COV19D-Competition}. ( Last accessed on March, 17{$^{th}$} 2024).}
\end{itemize}

This paper is organized in following way:  Section \ref{S:1} describes our proposed approaches for Covid-19 Detection and Covid-19 Domain Adaptation Detection. The experiments and results are detailed in Section  \ref{S:2}. Finally, we conclude our paper in Section \ref{S:3}.

\section{Our Approaches}
\label{S:1}

Our approach is tailored to capitalize on region of interest segmentation, specifically lung segmentation, and infection segmentation alongside input slices from CT scans. The objective is to develop a model proficient in discerning COVID-19 cases from non-COVID-19 cases. We evaluate three baseline architectures: Customized Hybrid-DeCoVNet \cite{bougourzi2023deep}, 3D-ResNet-18, and 3D-ResNet-50 \cite{hara2018can}.




\subsection{Data Preprocessing}

The objective of this phase is to eliminate slices that do not exhibit any lung structures and to identify lung features in the remaining slices. Following our previous approach in the 2nd COV19D competition  and  3rd COV19D competition \cite{bougourzi2023cnr, bougourzi2023deep}, ResneXt-50 \cite{Resnext} is used to filter the CT slices that does not show lung regions, to concentrate only on the slices that may have infection.

\subsection{Customized Hybrid-DeCoVNet}
In this challenge, we adopted our proposed Customized Hybrid-DeCoVNet which were proposed for Covid-19 severity prediction, to perform Covid-19 recognition in this challenge. The first modification is by considering the input slice, their region of interest segmentation and the infection segmentation as input, these three images are concatenated to form 3 channels. For segmenting the lung and the infection we used the PDAtt-Unet \cite{bougourzi2023pdatt} for segmenting the infection and the lung simultaneously. PDAtt-Unet is trained using three datasets Segmentation dataset nr. 2 \cite{COVID-19-Dataset}, COVID-19 CT segmentation \cite{COVID-19-Dataset} and CC-CCII \cite{liu2020kiseg}, each dataset is divided into 70\%-30\% as training and validation splits, then PDAtt-Unet is trained on the ensemble of 70\% of the three datasets and evaluated on the ensemble of the their 30\%. 

\begin{figure*}
\centering
    \includegraphics[width = 6in, height = 2.5in]{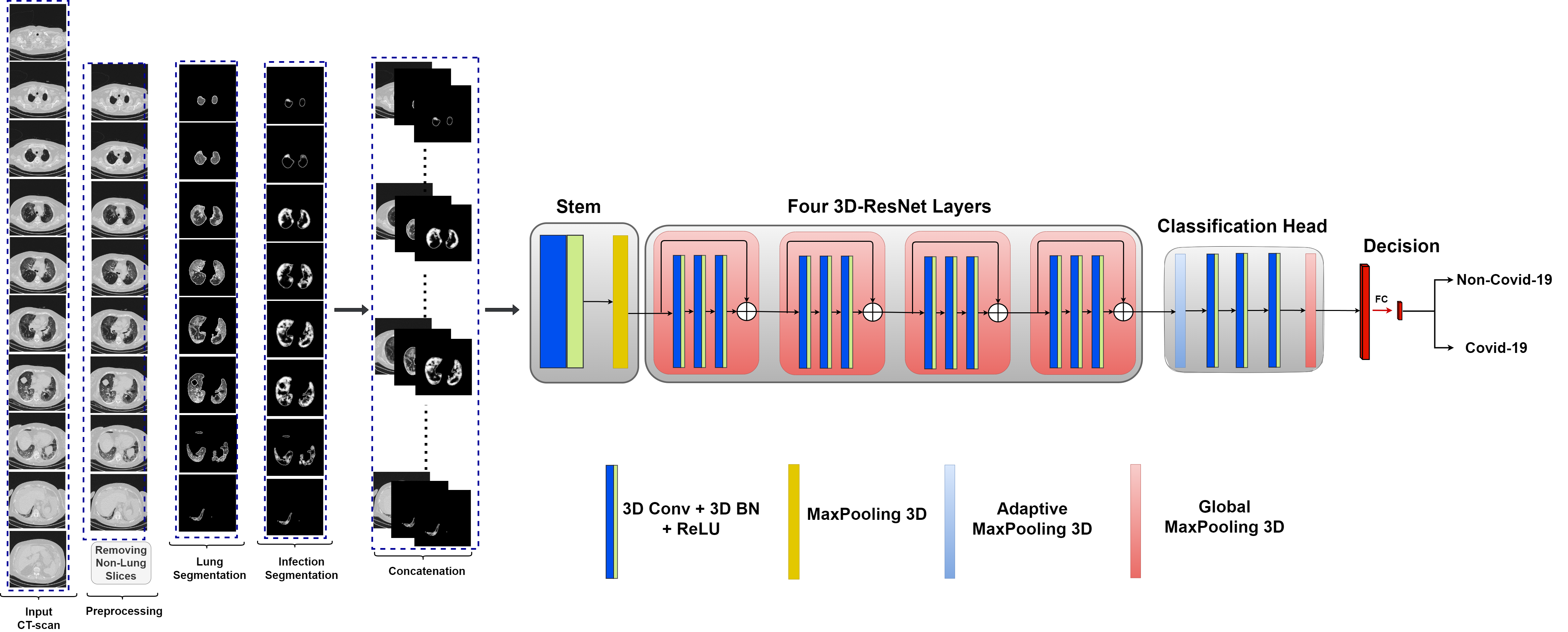} 
    \caption{Customized Hybrid-DeCoVNet Approach.}
    \label{fig:approach3}
\end{figure*}

As illustrated in Figure \ref{fig:approach3}, Customized Hybrid-DeCoVNet comprises of four components. First, the three images depicting the input slice, the segmented lung and segmented infection are merged into a three-channel image. This is performed for every slice of the input CT-scans, then all of these merged 3 channels images are concatenated. For a CT-scan of $N$ slice this will produce $S = 224\times224\times3\times N$. Since the number of slices is different from one CT-scan to another, S is resized into a fixed size of $224\times224\times3\times64$.   This resulting volume is fed into the Stem block, which is a 3D convolutional layer with a kernel of size (7, 7, 5) for height, width, and depth, respectively. The Stem block transforms the two input channels into 16 channels and is followed by Batch Normalization Layer (BN) and ReLU activation function. The second block of Customized Hybrid-DeCoVNet consists of four 3D-Resnet layers, which expand the channels to 64, 128, 256, and 512, respectively. The Classification Head comprises of 3D Adaptive MaxPooling, three 3D convolutional layers, and 3D Global MaxPooling. The output of the Classification Head is flattened into a single-channel deep feature map and fed into the Decision Head, which consists of one FC layer that has two outputs (Non-Covid-19 and Covid-19). Our proposed architecture is designed to enhance the performance of Covid-19 Prediction. It should be noted that Customized Hybrid-DeCoVNet does not have any pretrained weight in contrast with 3D-Resnet-18 and 3D-Resnet-50.

\subsection{3D-Resnet-18 and 3D-Resnet-50}

In addition to our proposed Customized Hybrid-DeCoVNet, we evaluated the use of pretrained 3D CNN architectures. To this end, we used the pretrained 3D-Resnet-18 and 3D-Resnet-50 from  \cite{hara2018can}. These pretrained models were trained for action recognition from 3D-video. For 3D-Resnet-18, the pretrained weights was trained on the ensemble of Kinetics-700 and Moments in Time datasets. While, 3D-Resnet-50 was trained on the ensemble of Kinetics-700, Moments in Time, STAIR-Actions datasets. To adopt these models for Covid-19 recognition, we changed the decision layer to give 2 output which corresponds to Non-Covid-19 and Covid-19 classes.

\section{Experiments and Results}
\label{S:2}

\subsection{The COV19-CT-DB Database}
In this competition, the COVID19-CT-Database (COV19-CT-DB) \cite{kollias2023ai,arsenos2023data,kollias2023deep,kollias2022ai,arsenos2022large,kollias2021mia,kollias2020deep,kollias2020transparent} is used for two sub challenges, which are Covid-19 Detection Challenge and Covid-19 Domain Adaptation Challenge. In Covid-19 Detection Challenge, many CT scans have been aggregated, each one of which has been manually annotated in terms of Covid-19 and non-Covid-19 categories. The resulting dataset is split into training, validation and test partitions. The provided training and validation partitions for developing the approach are summarized in Table \ref{tab:datasev}.

In the second challenge, Covid-19 Domain Adaptation Challenge, CT scans have been aggregated from various hospitals and medical centres. Each CT scan has been manually annotated with respect to Covid-19 and non-Covid-19 categories. The resulting dataset is split into training, validation and test partitions. Participants will be provided with a training set that consists of: i) the annotated data of the 1st Challenge which are aggregated from some hospitals and medical centres (case A); ii) a small number of annotated data and a larger number of non-annotated data (case B), all of which are aggregated from other hospitals and medical centres and their distribution is different from that of case A. Participants will be provided with a validation set that consists of a small number of annotated data of case B.

\begin{table}[H]
 \caption{Datasets summary of the 4th COV19D Competition. 494 Non annotated }
\label{tab:datasev}
 \begin{center}
\centering
\begin{tabular}{|l|c|c||c|c|}

\hline
{\multirow{2}{*}    \textbf{Sub-Competition}}    & \multicolumn{2}{|c|}{\textbf{Train}}& \multicolumn{2}{|c|}{\textbf{Validation}} \\
\cline{2-5}

 &\textbf{Covid-19} &   \textbf{Non-Covid-19} &\textbf{Covid-19} &   \textbf{Non-Covid-19}  \\
\hline
Covid-19 Detection Challenge &703  &  655  & 170 & 156 \\\hline 
Covid-19 Domain Adaptation Challenge& 120  &  120  & 56 & 113\\\hline 
\end{tabular}
\end{center}
\end{table}

\subsection{Experimental Setup}
We utilized the Pytorch Library and four NVIDIA GPU Device GeForce TITAN RTX 16 GB for Deep Learning training and testing.  The batch size of 16 CT-scan volumes was used to train the Customized Hybrid-DeCoVNet and 3D-Resnet-18 architectures for 80 epochs. While, a batch size of 8 is used to train 3D-Resnet-50 for 40 epochs. Warm up Cosine learning rate Schedule is adopted with initial learning rate of 0.0001. The following data augmentations are used for training and testing augmentation approach: random rotation with an angle between -40$^{\circ}$ to 40$^{\circ}$, vertical and horizontal flipping with a probability of 20\% for each, Multiplicative Noise, Random Brightness, Random Brightness Contrast, Random Contrast, and Random Grid Shuffle .

\subsection{Results}

\subsubsection{Results of the first evaluation scenario}
\label{sec:vlres}
In this part, we used the training data of Covid-19 Detection Challenge (first challenge) and the validation data to train and save the best model on the validation data after each epoch, this two splits will be denoted as Train1 and Val1. We also used the training and validation data of the Covid-19 Domain Adaptation Challenge (the second challenge) to evaluate the performance of our approach in unseen data, these two splits will be denoted as Train2 and Val2. Table \ref{tab:res1} summarizes the obtained results. From these results, it is noticed that the performance on the Train2 and Val2 splits decreased compared with the results on Val1, this is due to the change of data domain. However, the drop in results in not too big, this shows that our approach can achieve a good result. On the other hand, the ensembling approach achieves better performance on Train2 compared with the single architectures.

\begin{table}[H]
 \caption{Results of the first evaluation scenario without testing augmentation}
\label{tab:res1}
 \begin{center}
\centering
\begin{tabular}{|l|c|c||c|c||c|c|}

\hline
{\multirow{2}{*}    \textbf{Architecture}}    & \multicolumn{2}{|c|}{\textbf{Val1}}& \multicolumn{2}{|c|}{\textbf{Train2}}& \multicolumn{2}{|c|}{\textbf{Val2}} \\
\cline{2-7}

 &\textbf{Accuracy} &   \textbf{F1-score} &\textbf{Accuracy} &   \textbf{F1-score} &\textbf{Accuracy} &   \textbf{F1-score}  \\
\hline
 Customized Hybrid-DeCoVNet& 92.33 &  92.33  & 83.75  &  83.72  & 82.58  & 82.09 \\\hline 
3D-Resnet-18&  91.41  &  91.41    & 82.08  & 81.88   &  83.70& 82.59  \\\hline 
3D-Resnet-50& 91.41  &  91.40   & 84.58  & 84.58   & 83.70  &  82.59 \\\hline
Ensemble& 91.10  &   91.10  & 85 &  84.93 & 83.70 & 82.59 \\\hline

\end{tabular}
\end{center}
\end{table}

\subsubsection{Results of the second evaluation scenario}
\label{sec:tsres}

In the second evaluation scenario, we combined the training data of Covid-19 Detection and Covid-19 Domain Adaptation  challenges (Train1+Train2) in order to compare the performance of the three backbones in the scenario where the training data is augmented. The obtained results are summarized in Table \ref{tab:res3}, in which, Val1 and Val2 correspond to the validation of the first and the second challenge, respectively (correspond to the same splits used in Table \ref{tab:res1}). By comparing the results of Tables \ref{tab:res1} and \ref{tab:res3}, it is noticed that augmenting the training data improve the performance of the three backbones, especially 3D-Resnet-18. 

Table \ref{tab:res4} depicts the results of using testing augmentation, in which, each CT-scan is augmented ten times and the CT-scan prediction corresponds to the average probabilities of the prediction of the ten augmentations. Compared with the results of Table \ref{tab:res3}, using testing augmentation further improves the performance.

\begin{table}[H]
 \caption{Results of the second evaluation scenario without testing augmentation}
\label{tab:res3}
 \begin{center}
\centering
\begin{tabular}{|l|c|c||c|c|}

\hline
{\multirow{2}{*}    \textbf{Architecture}}    & \multicolumn{2}{|c|}{\textbf{Val1}}& \multicolumn{2}{|c|}{\textbf{Val2}} \\
\cline{2-5}

 &\textbf{Accuracy} &   \textbf{F1-score} &\textbf{Accuracy} &   \textbf{F1-score}  \\
\hline
 Customized Hybrid-DeCoVNet & 92.33   &  92.33    &  83.14  & 80.64  \\\hline 
3D-Resnet-18&  92.33  &  92.32    &  87.07  &  85.60 \\\hline 
3D-Resnet-50&  92.63  &  92.60   &  84.26  & 82.92  \\\hline 

\end{tabular}
\end{center}
\end{table}
\begin{table}[H]
 \caption{Results of the second evaluation scenario with testing augmentation}
\label{tab:res4}
 \begin{center}
\centering
\begin{tabular}{|l|c|c||c|c|}

\hline
{\multirow{2}{*}    \textbf{Architecture}}    & \multicolumn{2}{|c|}{\textbf{Val1}}& \multicolumn{2}{|c|}{\textbf{Val2}} \\
\cline{2-5}

 &\textbf{Accuracy} &   \textbf{F1-score} &\textbf{Accuracy} &   \textbf{F1-score}  \\
\hline
 Customized Hybrid-DeCoVNet & 91.41  &  91.40   &  84.83 & 83.23  \\\hline 
3D-Resnet-18& 92.33   & 92.33    &  88.76 & 87.52 \\\hline 
3D-Resnet-50&  92.33  &   92.33  & 85.39  & 84.34 \\\hline 
Ensemble& 92.33   &  92.33   &  88.20 & 87.14 \\\hline 
\end{tabular}
\end{center}
\end{table}

\subsection{Comparison with the Baseline}

Table \ref{tab:comp}, depicts the comparison with the baseline results from \cite{kollias2024domain}. The comparison of our approach and the baseline results shows that our approach achieved better performance on both challenges. In more details, our approach achieved better performance than the baseline approach by 14.33\% in terms of F1-score for Covid-19 Detection Challenge. Similarly for Covid-19 Domain Adaptation Challenge, our approach achieved better performance than the baseline by  14.52\% in terms of F1-score.

\begin{table}[H]
 \caption{Results Comparison with the baseline}
\label{tab:comp}
 \begin{center}
\centering
\begin{tabular}{|l|c|c||c|c|}

\hline
{\multirow{2}{*}    \textbf{Architecture}}    & \multicolumn{2}{|c|}{\textbf{sub-challenge 1}}& \multicolumn{2}{|c|}{\textbf{sub-challenge 2}} \\
\cline{2-5}

 &\textbf{Accuracy} &   \textbf{F1-score} &\textbf{Accuracy} &   \textbf{F1-score}  \\
\hline
 Baseline & -  &   78   &  - &   73 \\\hline
 Customized Hybrid-DeCoVNet & 91.41  &  91.40   &  84.83 & 83.23  \\\hline 
3D-Resnet-18& 92.33   & 92.33    &  88.76 & 87.52 \\\hline 
3D-Resnet-50&  92.33  &   92.33  & 85.39  & 84.34 \\\hline 
Ensemble& 92.33   &  92.33   &  88.20 & 87.14 \\\hline 
\end{tabular}
\end{center}
\end{table}

\section{Conclusion}
\label{S:3}
In this paper, we introduced a new approach for addressing the Covid-19 Detection and Covid-19 Domain Adaptation Challenges. Our approach primarily leveraged lung segmentation and Covid-19 infection segmentation through the utilization of state-of-the-art CNN-based segmentation architecture, namely PDAtt-Unet. This architecture enables simultaneous segmentation of lung regions and infections. Rather than feeding individual input slices to the training network, we concatenated the input slice (grayscale) with the segmented lung and infection, resulting in three input channels akin to color channels.

Moreover, we employed three distinct 3D CNN backbones to train Covid-19 recognition for both challenges: Customized Hybrid-DeCoVNet, as well as pretrained 3D-Resnet-19 and 3D-Resnet-50 models. To further enhance performance, we investigated ensemble approaches and testing augmentation techniques. Comparative analysis against baseline results demonstrates the significant efficiency of our proposed approach, exhibiting a substantial margin in terms of F1-score (14\%).

Our findings underscore the effectiveness of our methodology in addressing the complexities of Covid-19 detection and domain adaptation challenges. By integrating cutting-edge segmentation architectures and leveraging ensemble strategies, our approach demonstrates promising advancements in accurately identifying Covid-19 infections, thus contributing to the ongoing efforts in combatting Covid-19 pandemic and the future ones.




\end{document}